\documentclass[final,3p,onecolumn, times]{elsarticle}
\usepackage{graphicx}
\usepackage{subfig}
\usepackage{amsmath}
\usepackage{balance}
\usepackage{amssymb}
\usepackage{caption}
\usepackage{multirow}
\usepackage{booktabs}
\usepackage{color}
\usepackage[bottom]{footmisc}
\usepackage{epstopdf}
\usepackage{threeparttable}
\usepackage{longtable}
\usepackage{mathtools, cuted}
\usepackage{lipsum, color}
\newcommand{\parallelsum}{\mathbin{\!/\mkern-5mu/\!}}
\biboptions{square,sort&compress}

\begin{document}

\author[kth]{Song Lu\corref{song}}
\cortext[song]{Corresponding author. Tel: +46 8 7906215. Email: songlu@kth.se or lusommmg@hotmail.com (Song Lu)}
\author[kth]{John \AA gren}
\author[kth,uppsala,hungary]{Levente Vitos}

\address[kth]{Department of Materials Science and Engineering, Royal Institute of Technology, Stockholm SE-100 44, Sweden}%
\address[uppsala]{Department of Physics and Astronomy, Division of Materials Theory, Uppsala University, Box 516, SE-751210, Uppsala, Sweden}%
\address[hungary]{Research Institute for Solid State Physics and Optics, Wigner Research Centre for Physics, Budapest H-1525, P.O. Box 49, Hungary}%

\title{Ab initio study of energetics and structures of heterophase interfaces: from coherent to semicoherent interfaces}

\begin{abstract}
Density functional theory calculations have been performed to study the structures and energetics of coherent and semicoherent TiC/Fe interfaces.
A systematic method for determining the interfacial energy for the semicoherent interface with misfit dislocation network has been developed. The obtained interfacial energies are used to calculate the aspect ratios for the disc-like precipitates and a quantitative agreement with the experimental results is reached.  Based on the obtained interfacial energies  and atomic structure details, we propose models for describing the evolution of the interfacial energy with respect to the size of TiC precipitate for heterogeneous nucleation on an edge dislocation, shedding light on the thermodynamics of precipitate nucleation and growth. The present method can be easily applied to any heterophase interfaces between metals and oxides/carbides/nitrides.  
\end{abstract}

\maketitle

\section{Introduction}

Heterophase interfaces between metals and oxides/carbides/nitrides are abundant in materials  and can have a decisive role for the mechanical, chemical, and functional properties.  A fundamental understanding of the structure and energetics of the interfaces is an important step toward quantitative predictions of their impact on microstructure and properties of the material. 

In both austenitic and ferritic steels, transition metal carbides/nitrides (MX, M=Mo, Nb, Ti, Zr, Hf, V, etc. and X= C, N) are widely used precipitates to adjust the mechanical properties. In particular, in low carbon steels, such as high-strength low-alloy (HSLA) steels, homogeneously distributed nano-sized MX precipitates significantly increase the yield stress. In order to exploit the benefits of the MX precipitates in steels, i.e., precipitation strengthening, grain refinement, and enhancing resistance to hydrogen embrittlement  by trapping hydrogen \cite{NAGAO2014244,Davide2016,Wei2006}, etc.,  a detailed characterization of the structures and energetics of the phase interfaces is required. For example, prevention of precipitate coarsening by Ostwald ripening during subsequent processing is crucial to realizing a sufficient precipitation hardening effect  in the final product. One of the most fundamental factors to control is the interfacial energy which acts as the driving force for precipitate coarsening\cite{Jang2013}.  Therefore, information about the effects of alloying elements on the interfacial energy is of significant importance in alloys for composition design and optimization, e.g. to reveal the effects of Mo in retarding carbide precipitate coarsening in Ti-Mo containing steels. \cite{Jang2012208,Jang2013, MUKHERJEE20132521, MUKHERJEE2017621}


To date, there are no accurate experimental methods for measuring/estimating interfacial energy. The common procedure for measuring the interfacial energy is through adjusting the interfacial energy in the coarsening model (e.g., Ostwald ripening) to fit the observed evolution of precipitate size. \cite{2003420, ZHANG2017166} The resultant interfacial energies are associated with great uncertainties. When it concerns the early stage of precipitation, down to the size of several nanometers, or in magnetic systems like steels, characterizing the structures and energetics of interfaces becomes a seriously challenging task. 
On the other hand, as one of the most important physical parameters
in various thermodynamic models, \cite{DESCHAMPS1998293, Maugis2008,Jang2013,ZHANG2017166,Yang2002184, Jang2013} the interfacial energy is often treated more like a fitting parameter that ensures the match between experiment and modeling, e.g., to reproduce the observed evolution of the density and size of precipitates in different stages, i.e., nucleation, growth and coarsening. However, the variation of interfacial energy with respect to factors such as chemistry, the size of precipitates (coherent/semicoherent/noncoherent interfaces), the type of nucleation (homogeneously in bulk or heterogeneously on dislocations), and temperature, etc., are usually ignored or treated by {\it ad~hoc} parameters, \cite{Maugis2008} mainly due to the lack of accurate interfacial energy database, both from theoretical calculations and measurements. Further, accurate Density Functional Theory (DFT) calculations are normally restrained to the fully coherent interface, \cite{Jung2008, Jung2006,Park2013,Jang2012208,Davide2016} and only in a very few cases, semicoherent interfaces with significantly large mismatches were studied. For example, the MgO/Cu interface with $\sim$14\% mismatch was studied by direct DFT calculations. \cite{Benedek2000} With such a large lattice mismatch, it should be seen as an incoherent interface with a geometrical misfit-dislocation network rather than as a regular semicoherent interface. \cite{Zhang2014} 
Modeling semicoherent interface with small to medium lattice misfit containing interface dislocations requires thousands of atoms and are usually considered beyond the capability of nowadays DFT calculations. There are attempts to develop semidirect approaches to include strain energy contribution and also the misfit dislocation core part based on ab initio simulations for small supercells, \cite{Benedek2002,Benedek2004,Jung2010} but the accuracy of the resultant interface energetics are however difficult to assess. Furthermore, these approaches do not provide deep insights of the atomic/magnetic/chemical structures of the interfaces. An alternative approach to access the structure and energetics of semicoherent interfaces was developed within a  Peierls-Nabarro framework, in which ab initio data for the chemical interactions across the interface (the so-called $\gamma$ surface) are combined with a continuum description to account for the elastic distortions. The approach has been applied to the metal-oxide (Ag/MgO\cite{Zhang2007}  and Al/MgO\cite{zhang2008}),the metal-nitride/carbides,\cite{Johansson2005,Fors2010,FORS2010550}  and the metal-metal NiAl/Mo\cite{MEDVEDEVA2004675} interfaces. However, there is an ambiguity in the accuracy of the obtained interfacial energies  because the atomic configuration around misfit dislocations is not reproduced accurately, which becomes significant in the region where dislocations cross each other. The elastic contribution to the total interfacial energy is limited only to the interface layer.\cite{FORS2010550}  Another weakness of  this technique is that it is not appropriate for studying interfaces with alloying element segregation or vacancies.




 In the present work, we take the Fe/TiC interface as a model system to explore how one can obtain the interfacial energy for the semicoherent interfaces with misfit dislocations on the DFT level. The rest of this paper is arranged as follows. We present the calculation details and methodologies for modeling coherent and semicoherent interfaces in Section 2. The obtained energetics and structural details of the coherent and semicoherent interfaces are presented in Section 3. In Section 4,  first we discuss the development of morphology with respect to the size and coherency state of precipitates, then we propose two models for the evolution of the interfacial energy, which correspond to different nucleation positions for TiC precipitates with the existence of an edge dislocation. We further show that vacancy at the interface is intrinsic because of the negative formation energy at the intersection of misfit dislocations. 

\section{Methods and models}

The Vienna ab initio simulation package (VASP) is used to perform the electronic structural calculations.~\cite{PhysRevB.54.11169,kresse1996efficiency,kresse1993ab} The generalized gradient approximation (GGA) as parameterized by Perdew and Wang \cite{perdew1992pair} with the projector augmented waves (PAW) method~\cite{blochl1994projector, kresse1999ultrasoft} is adopted for the exchange correlation potentials of Fe, Ti, and C. Spin-polarized calculations are performed.
The calculated lattice parameters of body-centered cubic (bcc) Fe and TiC with NaCl structure are 2.836 \AA~and 4.331  \AA~, respectively. These values are in good agreement with previous theoretic and experimental data. \cite{Wei2006, Fors2010} The TiC precipitates have a disc-like morphology, the aspect ratio ($d/h$, $d$ diameter and $h$ height) of which changes with the  size of the precipitate. \cite{Wei2006}  
The experimental observed  Baker-Nutting (B-N) orientation relationship (OR), i.e., (100)Fe//(100)TiC and [100]Fe//[110]TiC, is used in the present study for the broad semicoherent interface.  It is generally expected that small TiC precipitates form coherent interface with bcc Fe.  Misfit dislocation networks were observed on the broad interface when the particle diameter ($d$) is larger than about 42 \AA~ and the interface loses full coherency to release coherent elastic energy, i.e., becoming a semicoherent interface. \cite{wei2004}  

The coherent interface (denoted as 1$\times$1Fe/1$\times$1TiC) has a very large lattice mismatch as measured by $\delta=2(a_{\rm TiC}\sqrt{2}/2-a_{\rm Fe})/(a_{\rm TiC}\sqrt{2}/2+a_{\rm Fe})=7.78\%$. In order to obtain coherency, in our model the Fe lattice is stretched while TiC  lattice is fixed to the equilibrium value under the consideration that TiC is much stronger (larger bulk modulus). \cite{Fors2010} The convergence of coherent interfacial energy against the numbers of Fe and TiC layers is tested and 5 layers of Fe(001) and 5 layers of TiC(001) are enough to ensure converged results. The interfacial energy map for the coherent interface ($\gamma$-surface) is calculated by changing the relative position between the Fe and TiC lattices. The starting configuration (origin) is set as the interface where the Fe atoms sit on top of the interface C atoms (Fe-on-C, $(x,y)=(0,0)$) and the displacement vector is defined as ${\bf r}=x{\bf a}+y{\bf b}$, ($0\leq x,y\leq 0.5$). ${\bf a}= a_{\rm TiC}/2[110]$ and ${\bf b}=a_{\rm TiC}/2[-110]$ are the lateral lattice vectors of the coherent interface. The positions $(x,y)$ equal to $(0, 0.5)$ and $(0.5, 0.5)$ correspond to the Bridge and  Fe-on-Ti  configurations, respectively (Fig.1). The length of the supercell and the atomic positions along the $\bf c$ direction (perpendicular to the interface) are fully relaxed. The atomic forces are relaxed to less than 0.02 meV \AA$^{-1}$. A Monkhorst-Pack mesh of 10$\times$10$\times$3 k-points is adopted for the coherent interface calculations.

 For the preferred orientation relationship between TiC and Fe, the misfit dislocation network at the semicoherent interface is expected to be square and the periodicity $p$ of the dislocations is given by $p=Pa^{<110>}_{\rm TiC}=(P+1)a_{\rm Fe}$, where $a^{<110>}_{\rm TiC}=\sqrt{2}a_{\rm TiC}/2$, and $P$ is an integer. In the present case, the minimum $P$ that leads to the coincidence-site-lattice type of  periodicity is 12 (0.32\% mismatch). A proper supercell modeling this semicoherent interface (13$\times$13 Fe/12$\times$12 TiC) with, e.g., five Fe(100) and five TiC (100) layers along the direction perpendicular to the interface consists of a total of 2285 atoms, which is too expansive for direct ab initio calculations. 

Instead, we start with the so-called $\it 1D- semicoherent~interface$ (1D-SI), which is modeled by 13$\times$1 Fe units matching 12$\times$1 TiC units, Fig.\ref{fig1d}. In this case, the misfit strain is nearly totally released along  the $\bf{a}$ direction (with 0.32\% mismatch), while along $\bf{b}$ the interface maintains coherency  (with 7.78\% mismatch). 
Similarly to the $\gamma$-surface calculation for the coherent interface, the relative positions between the Fe and TiC lattices along  $\bf{b}$  in the 1D-SI are considered by rigid shifts of the two parts. The displacement vector is also given by ${\bf r}=x{\bf a}+y{\bf b}$, ($x=0$, $0\leq y\leq 0.5$). When $y=0$,  the relative position between interface Fe and C atoms changes  from the  Fe-on-C to the Bridge configurations, while for $y=0.5$, it goes gradually from the Bridge to the Fe-on-Ti positions, Fig.\ref{fig1d}.  By construction, this model describes sharp interfaces with parallel misfit dislocations with special dislocation core structures. Optimization of the length of lattice vector $\bf{c}$ and  atomic relaxation  along $\bf{a}$ and $\bf{c}$  are performed.   The Monkhorst-Pack k-points mesh of 2$\times$5$\times$3 is adopted for the calculations for the 1D-SIs. The convergence criteria for atomic forces is set again to  0.02 meV \AA$^{-1}$. Convergence test shows that 5 layers of Fe(001) are enough to  obtain the interfacial energy for the 1D-SI with an error less than 0.02 J m$^{-2}$.
For the calculation of the  interfacial energy, the reference Fe lattice (with energy $E_{\rm Fe}^{\rm 13\times1Fe/12\times1TiC}$ per atom) was calculated using a unit cell of body-centered cubic (bct) Fe at the same strain state as in the 1D interface.  Furthermore, we have calculated the energy for the strained Fe in the 1D interface model by an incremental method, i.e, $E_{\rm Fe}^{\rm 1D-13\times1Fe/12\times1TiC}$ from the total energy difference between the 1D interfaces with 9 and 5 layers of Fe along the $\bf c$ direction, respectively. The obtained energy differs from the unit cell calculation by less than 2 meV/atom.

The interfacial energies for the coherent and 1D semicoherent interfaces are calculated according to 
\begin{equation}\label{eq1}
\sigma=\frac{E_{\rm Fe/TiC}-NE_{\rm strained, Fe}-ME_{\rm TiC}}{2A},
\end{equation}
where $E_{\rm Fe/TiC}$ is the total energy of the supercell with two interfaces, and $E_{\rm strained, Fe}$ ( per Fe atom) is the reference energy of the Fe lattice under the same strain state as in the corresponding interface supercell. $E_{\rm TiC}$ (per TiC) is the total energy of the TiC carbide at the equilibrium state since in the present work we always distort the Fe lattice to match the TiC lattice. $N$ and $M$ are the corresponding numbers of lattice sites of Fe and TiC in the supercell, respectively.  $A$ is the interface area. In this way, the obtained interfacial energy is not volume dependent, and it is considered as the chemical part of total interfacial energy ($\sigma^{\rm chem.}$). On the other hand, the elastic energy contribution to the total interfacial energy in the coherent interface and 1D-SIs can be evaluated from the energy difference between the distorted Fe and the equilibrium bcc Fe, ($E_{\rm strained, Fe}-E^{\rm eq.}_{\rm Fe}$). This contribution is volume dependent. 

We emphasize that for the ideal semicoherent interfaces where the lattice mismatch strains are completely released by interface dislocations, the total interfacial energy should therefore be referred to the equilibrium bulk phases.



\section{Results}

\subsection{Coherent interfaces}

In Table 1 we present the calculated interfacial energies and structural/magnetic details for the coherent interfaces with different translation symmetries, i.e., Fe-on-C, Fe-on-Ti and bridge,  accompanied with available literature data. In general, the calculated interfacial energies are consistent with previous results, considering the different calculation details, e.g.,  atomic/volume/shape relaxations, underlying lattice, exchange-correlation functionals, reference states, etc.\cite{Park2013, Davide2016,Jung2008,Jang2012208,Fors2010}  Notice that atomic relaxation (along $\bf c$ direction) leads to significantly reduced interfacial energy, especially for the Fe-on-C and bridge configurations (by $\sim$ 37\% and 28\%),  compared to that for the Fe-on-Ti interface ($\sim$ 10\%). We emphasize the significantly different energetic and structural/magnetic details for different coherent interface configurations.  The Fe-on-C configuration has the lowest interfacial energy, and the Fe-on-Ti has the highest one. In agreement with previous observations, the interface C atoms are found to move towards interface Fe atoms while Ti atoms move away from the interface, which causes the so called layer buckling. The interface separations ($d_0$) which may be evaluated as the average of $z_{\rm 0}^{\rm Fe-C}$  and $z_{\rm 0}^{\rm Fe-Ti}$ do not follow the same sequence as the interfacial energies. $z_{\rm 0}^{\rm Fe-C}$  and $z_{\rm 0}^{\rm Fe-Ti}$ measures the $c$ corrdinate differences for the interfacial Fe/C and Fe/Ti atoms, respectively. The fact that $d_0$ is different for different interface configurations and that it is different from the averaging value  (1.75 \AA) of the interlayer distances for TiC and bct Fe emphasizes the importance of local chemical environment in determining the properties of interfaces. In Table \ref{coh}, we also list $\Delta z$  which measures the $\bf c$ coordinate difference between interface C and Ti atoms. It can be seen that the Bridge interface has the largest $\Delta z$  and the Fe-on-C interface has the lowest one, noticing that $\Delta z$ does not follow the interfacial energy sequence established for these three interfaces ($\sigma_{\rm Fe-on-C}<\sigma_{\rm Bridge}<\sigma_{\rm Fe-on-Ti}$). Large $\Delta z$ may indicate a higher atomic-level interface roughness. This micro-scale roughness may serve as useful information for understanding macro-scale phenomenon such as interface friction and galling.\cite{Vitos2006193}   The magnetic moments of the interface Fe atoms increase at the Fe-on-Ti interface, while decrease at the Fe-on-C and Bridge interfaces. The change of the local magnetic moment is strongly related to the interface separation for the different interface configurations. 

We emphasize that the interfacial energies in Table 1 represents only the chemical part of total coherent interfacial energy. They do not depend on volume. On the other hand, the strain energy contribution to the total interfacial energy in the coherent interface model depends linearly on volume, in the present case, on the  number of Fe layers when fixing the lateral lattice constants to those of TiC. This contribution is calculated to be $\sim$0.144 J m$^{-2}$ per Fe atom, which is in nice agreement with previous DFT results.\cite{Jang2013, Jang2012208} For instance, for the present supercell the Fe-on-C interfacial energy increases from 0.22 J m$^{-2}$ to $0.22+0.144\times5/2=0.58$ J m$^{-2}$ when the strain energy contribution is included (Table 2). 

\begin{figure}[tbh!]
	\centering
	\includegraphics[scale=0.3]{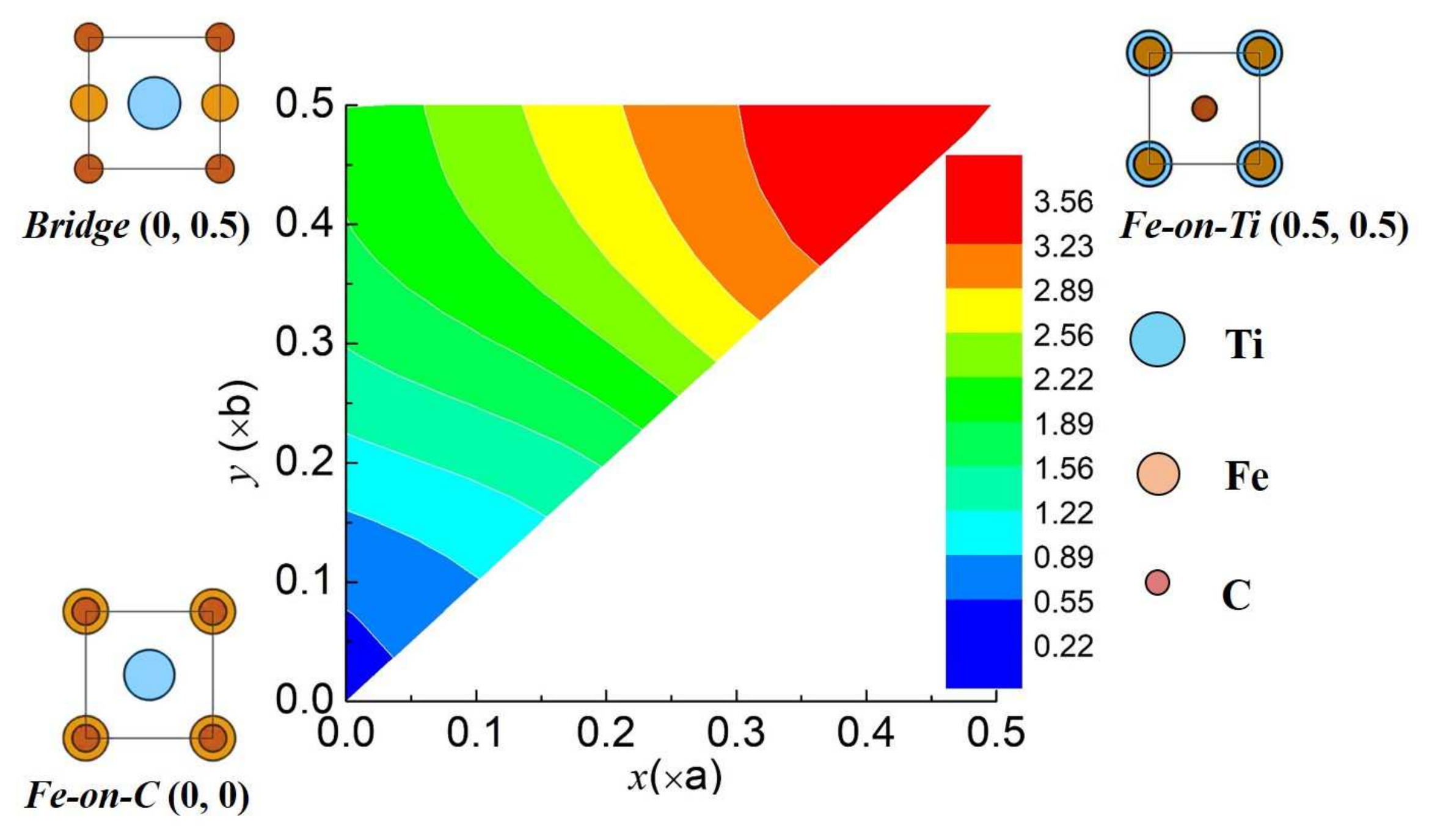}
	\caption{(Color online) The coherent interfacial energy map ($\gamma$-surface) for the Fe/TiC interface (in units of J m$^{-2}$). ${\bf a}= a_{\rm TiC}/2[110]$ and ${\bf b}=a_{\rm TiC}/2[-110]$ are the lattice vectors of the coherent interface. The maximum and minimum values are obtained for the Fe-on-Ti and Fe-on-C intefaces, respectively.}
\label{map}
\end{figure}

\begin{table*}[tb!]
\caption{The calculated interfacial energies ($\sigma_{\rm coh.}$, in units of J m$^{-2}$),  structural and magnetic details for the coherent interfaces. The relative changes for $d_1$ and $d_2$ compared to the reference bct Fe (001)  interlayer distance ($d_{\rm bct}^{\rm Fe}$) are listed in the parentheses (\%). $\Delta z$ (in units of \AA) is the difference between $z_{\rm 0}^{\rm Fe-Ti}$  and $z_{\rm 0}^{\rm Fe-C}$,  measuring the coordinate difference between the interface C and Ti atoms along the direction perpendicular to the interface. m$_{0,1,2}$ are the magnetic moments of Fe layers in the supercell, where the subscripts give the interlayer distance away from the interface Fe layer. m$_{\rm bct}^{\rm Fe}$ is the magnetic moment of the reference bct Fe which has the same strain state as in the coherent interface supercell.}
\label{coh}
\smallskip
\tiny
\begin{threeparttable}
\begin{tabular}{lllllllllllll}
\hline\hline
     &  $\sigma_{\rm coh.}$ & $z_{\rm 0}^{\rm Fe-C}$  & $z_{\rm 0}^{\rm Fe-Ti}$ & $\Delta z$ & d$_0$ &d$_{\rm 1}$ (\%) & d$_{\rm 2}$ (\%)& d$_{\rm bct}^{\rm Fe}$ & m$_{0}$ & m$_{1}$& m$_{2}$ & m$_{\rm bct}^{\rm Fe}$\\
\hline
Fe-on-C & 0.22, 0.35\tnote{$\dagger$}, 0.19\tnote{a}, 0.26\tnote{c}, 0.34$^\dagger$\tnote{d}, 0.39$^\dagger$\tnote{e}& 1.88 & 1.95 & 0.07, 0.06\tnote{a}, 0.07\tnote{b} &1.91&1.22 (-8.57) & 1.35 (1.80) & 1.33 & 2.16 & 2.49 & 2.54 & 2.56\\
Fe-on-Ti & 3.54, 3.92\tnote{$\dagger$}, 3.72$^\dagger$\tnote{e}& 2.65 & 2.80 & 0.15, 0.18\tnote{b}&2.73&1.21 (-9.02)& 1.35 (1.50 ) & 1.33 & 2.84 & 2.50 & 2.58  &2.56\\
Bridge & 1.78, 2.47\tnote{$\dagger$}, 2.53$^\dagger$\tnote{e}& 1.59& 1.96& 0.37, 0.35\tnote{b}&1.78& 1.23 (-7.37)& 1.36 (2.56) & 1.33 & 2.07 & 2.65 & 2.62 & 2.56\\
\hline\hline
\end{tabular}
\begin{tablenotes}
\item[a]Ref.~\cite{Park2013}. \item[b] Ref.~\cite{Davide2016}.
\item[c]Ref.\cite{Jung2008}. \item[d]Ref.\cite{Jang2012208}. 
\item[e]Ref.\cite{Fors2010}.
\item[$\dagger$] Interface distance relaxation without atomic relaxation. 
\end{tablenotes}
\end{threeparttable}
\end{table*}

\subsection{1D-semicoherent interface}

The calculated interfacial energies for the 1D-SIs as a function of $y$  are plotted in Fig.\ref{fig1de} (blue rhombi), and also summarized  in Table \ref{chem}. As expected from the coherent interfacial energy map, when most of the Fe atoms sit on top of C atoms ($y=0$), the obtained interfacial energy is the smallest, and the interfacial energy increases as the interface configuration getting close to the Bridge and Fe-on-Ti configurations, ($y\rightarrow 0.5$). 

The structure after relaxation is instructive for analyzing and understanding the results for semicoherent interfaces. For the case of $y=0$, we plot the atomic strains along $\bf{a}$ ($\varepsilon_{\rm Fe}^{\parallelsum}$) and the variation of interlayer distance at each Fe site  in Fig.\ref{fig1dstruc} (a) and (b), respectively. 
The Fe lattice away from dislocation core is found to expand to form coherent patches, meanwhile around dislocation core the Fe lattice is severely compressed, as what one expects with the existence of misfit dislocations.\cite{Zhang2014} In other words, the misfit dislocation induced strain $\varepsilon_{\rm Fe}^{\parallelsum}$  is modulated along $\bf{a}$, the integration of which is zero, as expected. The interface Fe layer experiences the strongest variation of $\varepsilon_{\rm Fe}^{\parallelsum}$ and the modulation of the strain decays fast from the interface Fe layer into the bulk, particularly at dislocation core region (see the insert plot in Fig.\ref{fig1dstruc} (a)). We emphasize that isotropic elastic theory or the classic continuum Peierls-Nabarro model are not able to describe strain near dislocation core and lead to diverging results. \cite{Zhang2014}
The decay of distortion can be more clearly seen from the Fe(001) interlayer distance in  Fig.\ref{fig1dstruc} (b). It shows that already from the second Fe layer, the interlayer distance converges approximately to the equilibrium value of the strained Fe lattice in the 1D-SI. The above results indicate that at  the semicoherent TiC/Fe interface the interface and misfit dislocation induced strains are primarily limited within 2-3 Fe layers next to the interface, which is in general agreement with previous molecular dynamic simulation results \cite{Johansson2005}. This conclusion is obtained for the 1D-SIs, which may also be true for the real 2D semicoherent interface.

 In Fig.\ref{fig1dstruc} (c) we show the longitudinal distance between neighboring interface Fe and C atoms for $y=0$ and $y=0.5$. The distance is scaled by $a^{<110>}_{\rm TiC}$. It shows that the interface Fe atoms away from dislocation core are relaxed to the positions with lower chemical interfacial energy as indicated by the coherent interfacial energy map (Fe-on-C for $y=0$ and Bridge for $y=0.5$). The atomic disregistry across the interface is plotted in Fig.\ref{fig1dstruc} (d), from which the half width of dislocation is calculated to be 2.38 \AA ~by the same definition as in the Peierls-Nabarro model. \cite{FORS2010550} This value is lower than the previous results (2.9-3.3 \AA~for Fe/VN interface \cite{FORS2010550}) obtained with the Peierls-Nabarro model. However, it confirms the importance of considering relaxation and elastic anisotropy  in both phases, which significantly decreases the interfacial energy and dislocation width. \cite{Johansson2005}

\begin{figure*}[tbh!]%
 \centering
\includegraphics[width=12cm]{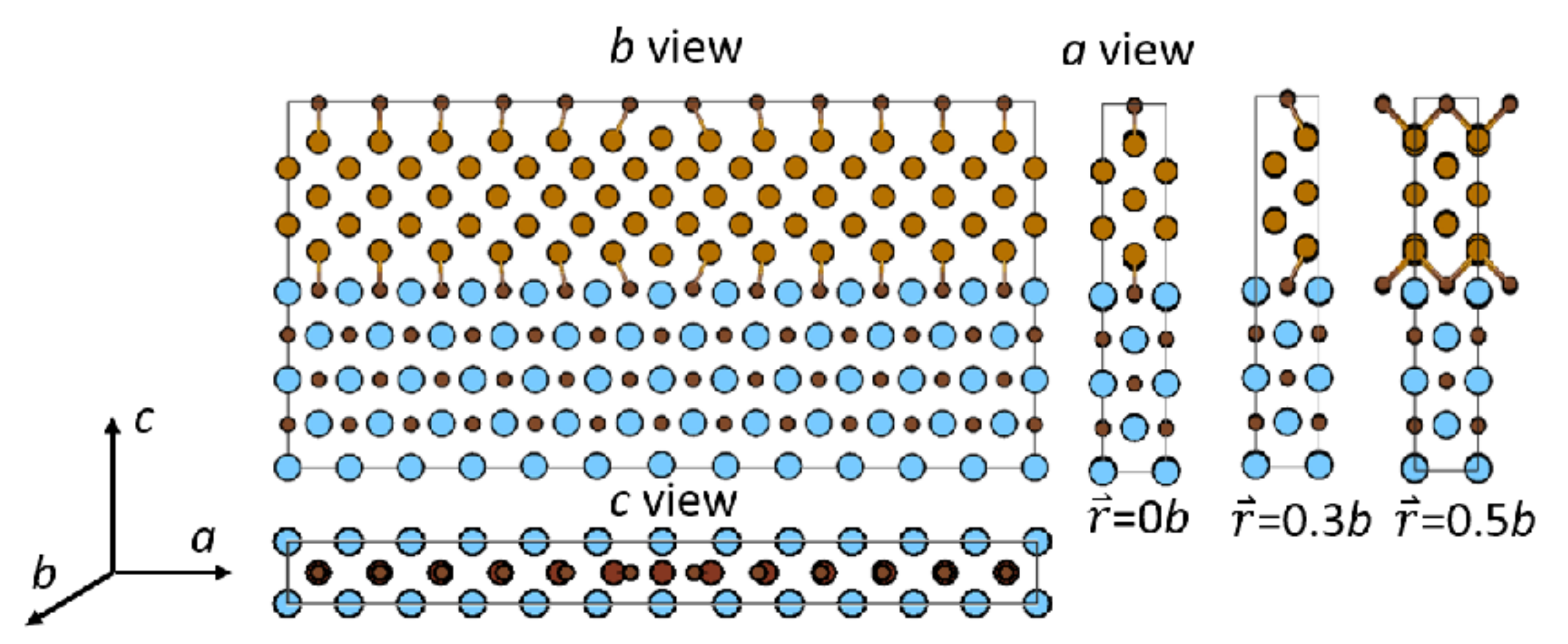}%
\caption{(Color online) (a) Schematic of the 1D-SI (13$\times$1Fe/12$\times$1TiC) after relaxation, viewed along $\bf a$, $\bf b$ and $\bf c$ directions. The rigid displacement between Fe and TiC lattice along $\bf b$ is described by ${\bf r}=x{\bf a}+y{\bf b}$, ($x=0, 0\leq y\leq0.5$). The bond lines between interfacial Fe and C atoms are guides for eyes.}%
\label{fig1d}%
\end{figure*}

\begin{figure}[tbh!]%
 \centering
\includegraphics[width=8cm]{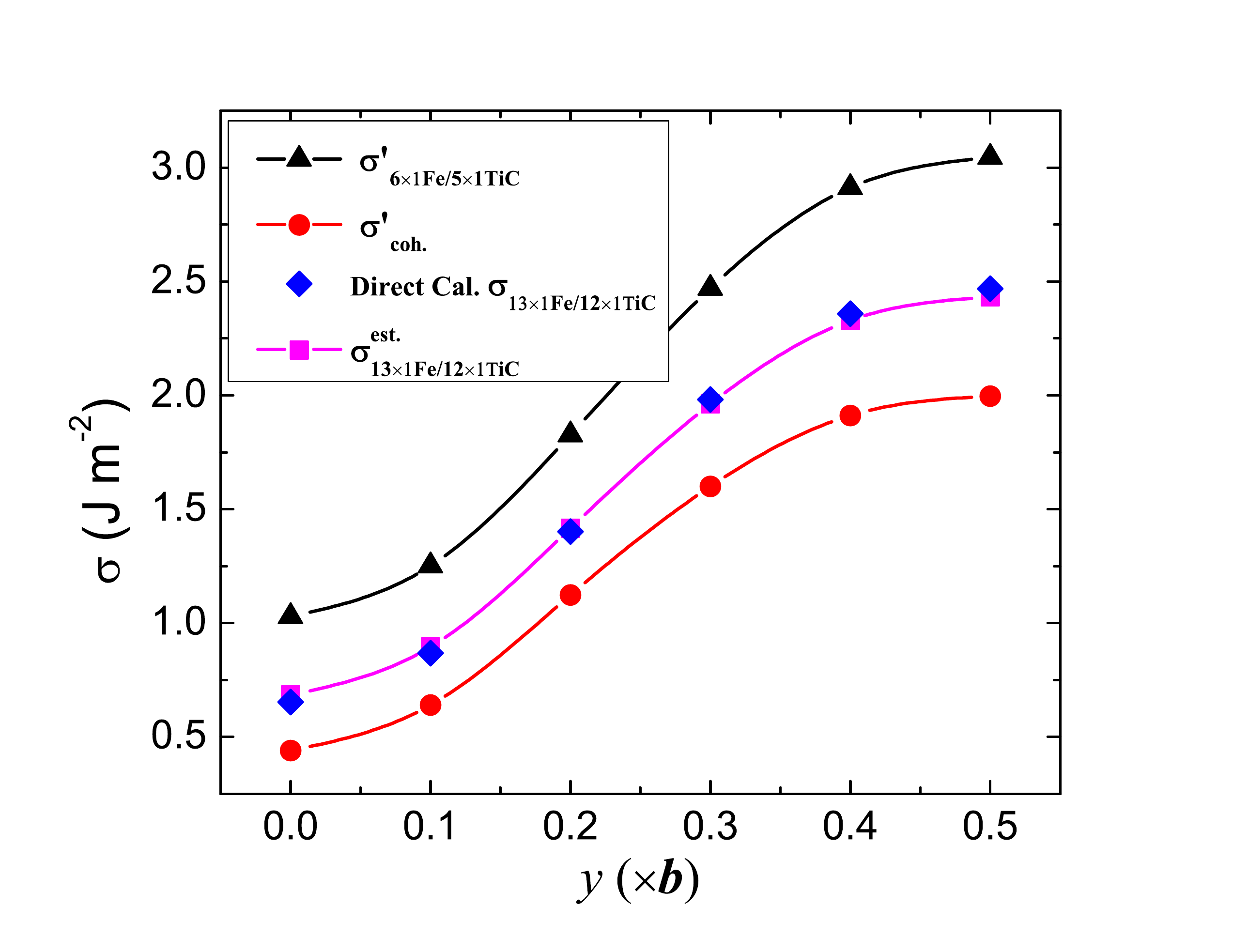}
\caption{(Color online) The calculated interfacial energies for the coherent  and semicoherent interfaces with respect to displacement along lattice vector $\bf{r}$. The reference energy for Fe is $E_{\rm Fe}^{1D-13\times1Fe/12\times1TiC}$ for all the coherent and semicoherent interfacial energies. See the text for notations.}%
\label{fig1de}%
\end{figure}

\begin{figure*}[tbh!]%
\centering
    \subfloat[]{{\includegraphics[width=8cm]{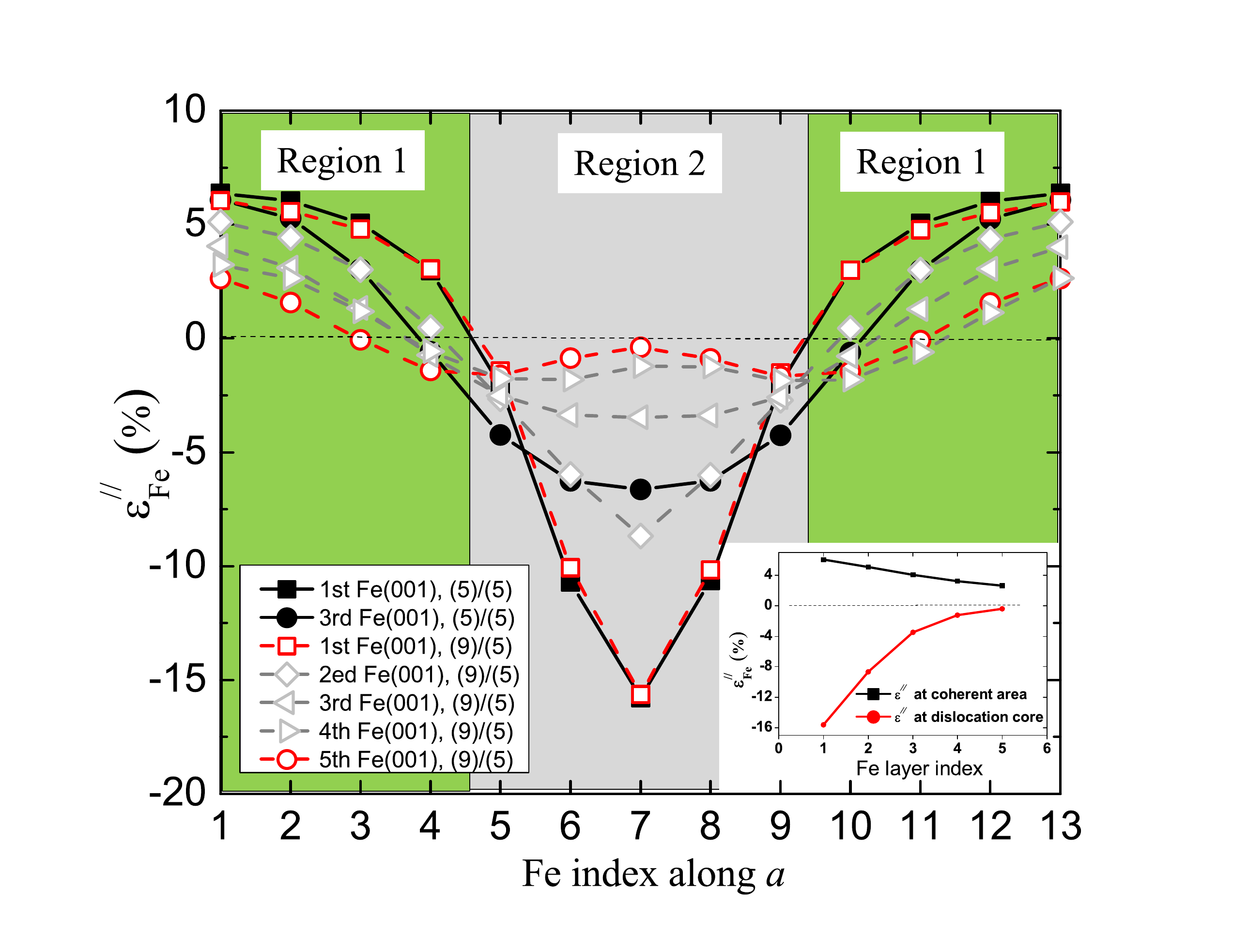} }}%
    \subfloat[]{{\includegraphics[width=8cm]{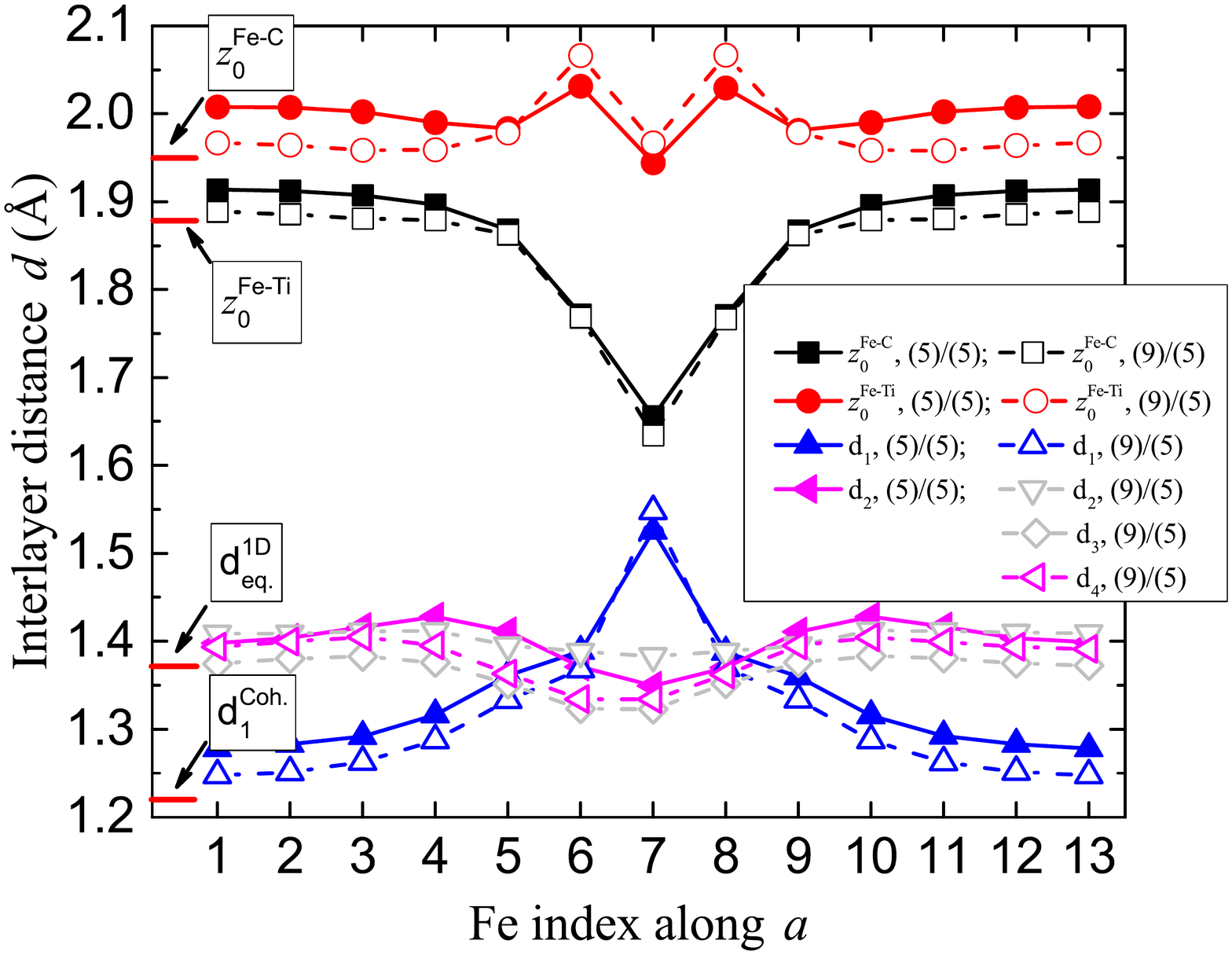} }}\\
    \subfloat[]{{\includegraphics[width=8cm]{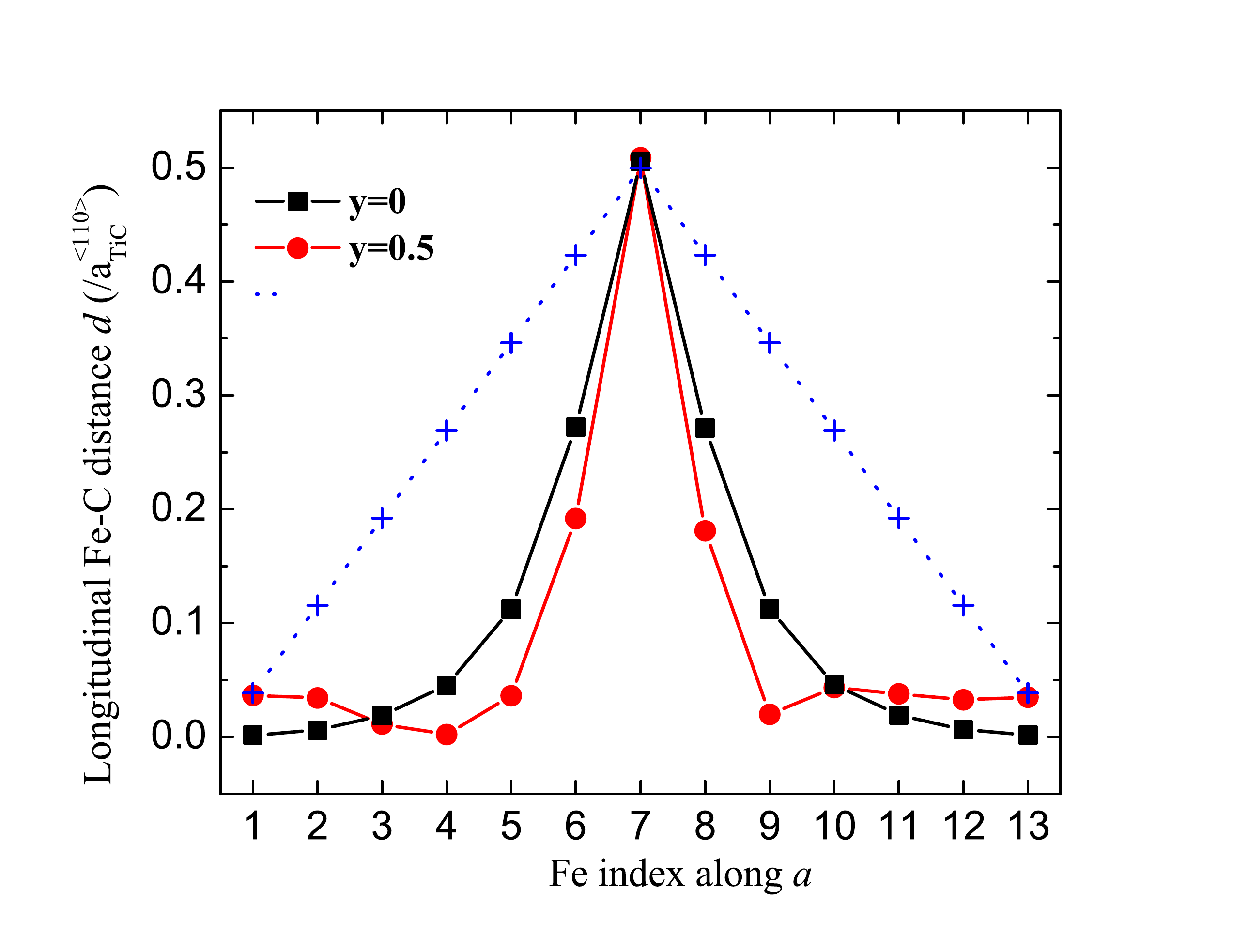} }}%
    \subfloat[]{{\includegraphics[width=8cm]{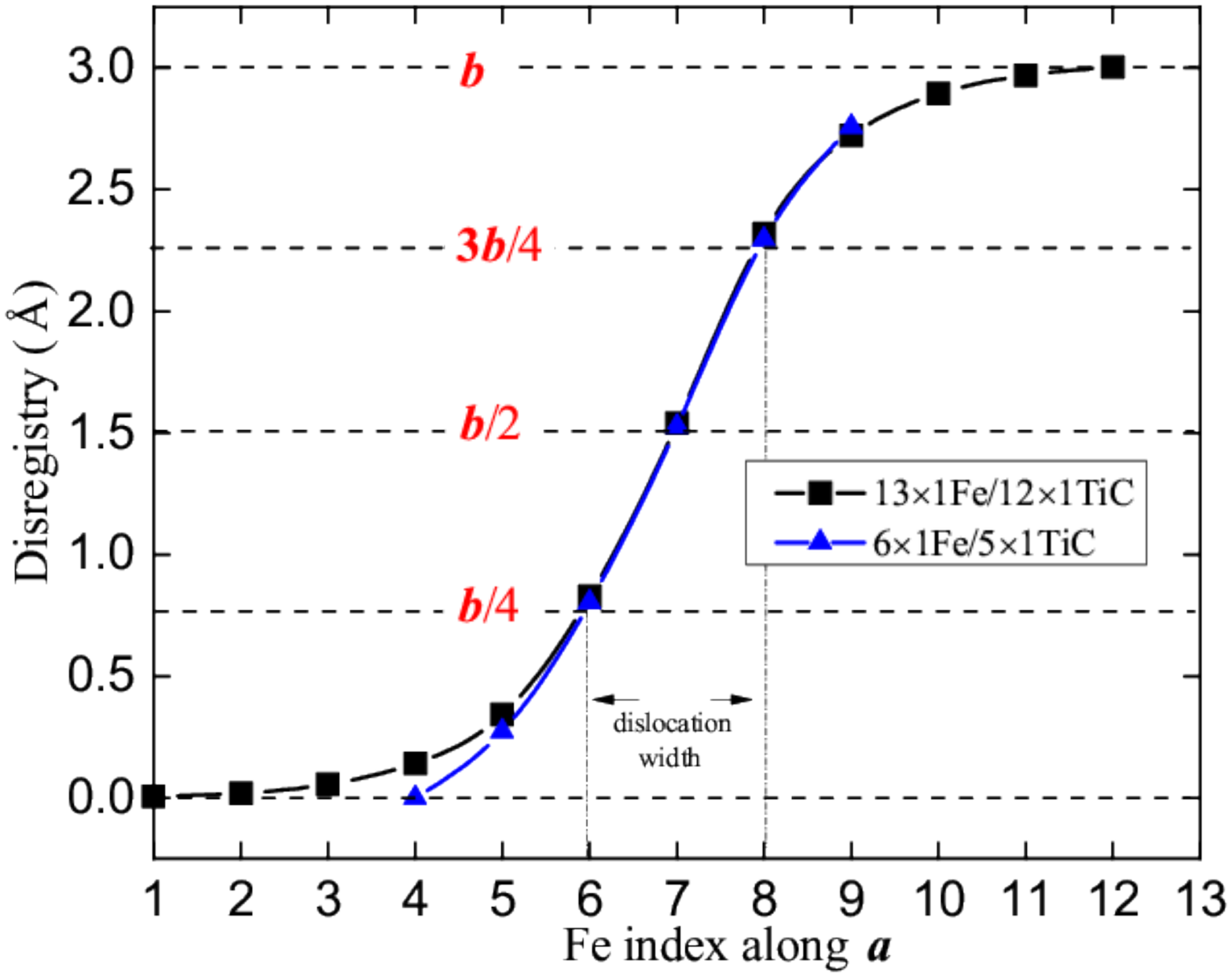} }}%
    \caption{(Color online) (a) The atomic strain along $\bf a$ ($\varepsilon_{\rm Fe}^{\parallelsum}$) in Fe(001)  layers  and (b) the variation of interlayer distance across the interface for the case $y=0$. The results for two 1D-SIs with 5 and 9 layers of Fe(100)  are plotted together for comparison ((5)/(5) and (9)/(5) for the Fe(5)/TiC(5) and Fe(9)/TiC(5) interfaces, respectively). $z^{\rm{0}}_{\rm{Fe-C}}$ and $z^{\rm{0}}_{\rm{Fe-Ti}}$ are the vertical distances between interfacial Fe-C and Fe-Ti atoms, respectively. $d_{1...5}$ are the Fe(001) interlayer distances with the subscripts indicating the distance away from the interface. The corresponding values in the case of coherent interface with Fe sitting on C are also marked in the plot. $d^{\rm{1D}}_{\rm{eq.}}$ is the equilibrium Fe interlayer distance in the 1D-SI. (c) The relative longitudinal distance between interface Fe and C atoms for $y=0$ (square) and $y=0.5$ (circle), in comparison with the ideal structure (plus symbol). (d) Atomic disregistry across the interface for the 1D 13$\times$1Fe/12$\times$1TiC and 6$\times$1Fe/5$\times$1TiC interfaces.} %
    \label{fig1dstruc}%
\end{figure*}

Next we make an attempt to identify and separate the core part of the misfit dislocation. Accordingly, the 13$\times$1Fe/12$\times$1TiC 1D-SI may be represented by $m$ coherent interface patches (region 1) and a smaller $(n+1)$ $\times$1Fe/$n\times$1TiC 1D-SI modeling the dislocation core (region 2). $m$ and $n$ are integers and $m+n=12$. With increasing $n$ close to 12, this division becomes more and more accurate. The estimated interfacial energy is then given by
\begin{equation}\label{intest1}
\sigma^{\rm est.}_{\rm 13\times1Fe/12\times1TiC}=f_1\sigma'_{\rm coh.}+f_2\sigma'_{\rm (n+1)\times1Fe/n\times1TiC},
\end{equation}
where $f_1=m/12$ and $f_2=n/12$ are the area fractions of the coherent and dislocation core regions, respectively. $\sigma'_{\rm coh.}$ and $\sigma'_{\rm (n+1)\times1Fe/n\times1TiC}$ are the interfacial energies for the coherent and  ($n$+1)$\times$1Fe/$n\times1$TiC 1D semicoherent interfaces calculated with $E_{\rm Fe}^{\rm 13\times1Fe/12\times1TiC}$ as the reference energy for Fe. The superscript ($\prime$) is used to highlight the fact that here for all interfacial energies we consider a specific (constant) reference energy of Fe. In other words, Eq.\ref{intest1} in fact expresses an estimation for the total energy of the supercell since for all three terms we adopt the same Fe reference energy.  
When we take $m=7$ and $n=5$ judged from above structure analysis, the estimated interfacial energies are in perfect agreement with the direct calculations, which justifies the above division of coherent and dislocation core areas. $\sigma'_{\rm coh.}$, $\sigma'_{\rm 6\times1Fe/5\times1TiC}$ and $\sigma^{\rm est.}_{\rm 13\times1Fe/12\times1TiC}$ are plotted in Fig.\ref{fig1d}(b) as a function of $y$.  

We have to emphasize that $\sigma^{\rm est.}$ in Eq.\ref{intest1} is volume dependent, more specifically, on the the number of Fe layers ($l_{\rm Fe}$) in the supercells,  via., 
\begin{equation}
\sigma'_{\rm coh.}=\sigma_{\rm coh.}^{\rm chem.}+\frac{ml_{\rm Fe}(E_{\rm Fe}^{\rm coh.}-E_{\rm Fe}^{\rm 13\times1Fe/12\times1TiC})}{2A_{\rm coh.}},
\end{equation}
 and  
\begin{equation}
\sigma'_{\rm (n+1)\times1Fe/n\times1TiC}=\sigma_{\rm \rm (n+1)\times1Fe/n\times1TiC}^{\rm chem.}+\frac{l_{\rm Fe}(n+1)(E_{\rm Fe}^{\rm \rm (n+1)\times1Fe/n\times1TiC}-E_{\rm Fe}^{\rm 13\times1Fe/12\times1TiC})}{2nA_{\rm coh.}}.
\end{equation}
$A_{\rm coh.}$ is the coherent  interface area. From above equations, an additional layer of Fe(001) (interface supercell containing 6 Fe(001) and 5 TiC(001)) is estimated to lead to an increment of $\sim$0.049 J m$^{-2}$ for the interfacial energy calculated by Eq.\ref{intest1}, (using data from Table 2). Therefore, the almost perfect agreement between the directly calculated interfacial energies and the estimated values by Eq.\ref{intest1} demonstrated in Fig.\ref{fig1de} implies that using the present setup (5 layers of Fe(001), two interfaces) we have properly taken into account the elastic energy contribution  in both coherent and the dislocation core areas, which is also in line with above structural analysis (Fig.\ref{fig1dstruc} (a) and (b)) and previous simulations~\cite{Johansson2005}.  

The atomic disregistry across the 6$\times$1Fe/5$\times1$TiC 1D-SI for $y=0$ is plotted in Fig.\ref{fig1dstruc} (d) and compared with that for the 13$\times$1Fe/12$\times$1TiC interface. The prefect agreement between each other again confirms that the dislocation core part has been adequately described by the truncated supercell (6$\times$1Fe/5$\times$1TiC ). 

\subsection{2D-semicoherent interface}

\begin{figure}[h]%
    \centering
    \subfloat[]{{\includegraphics[width=8cm]{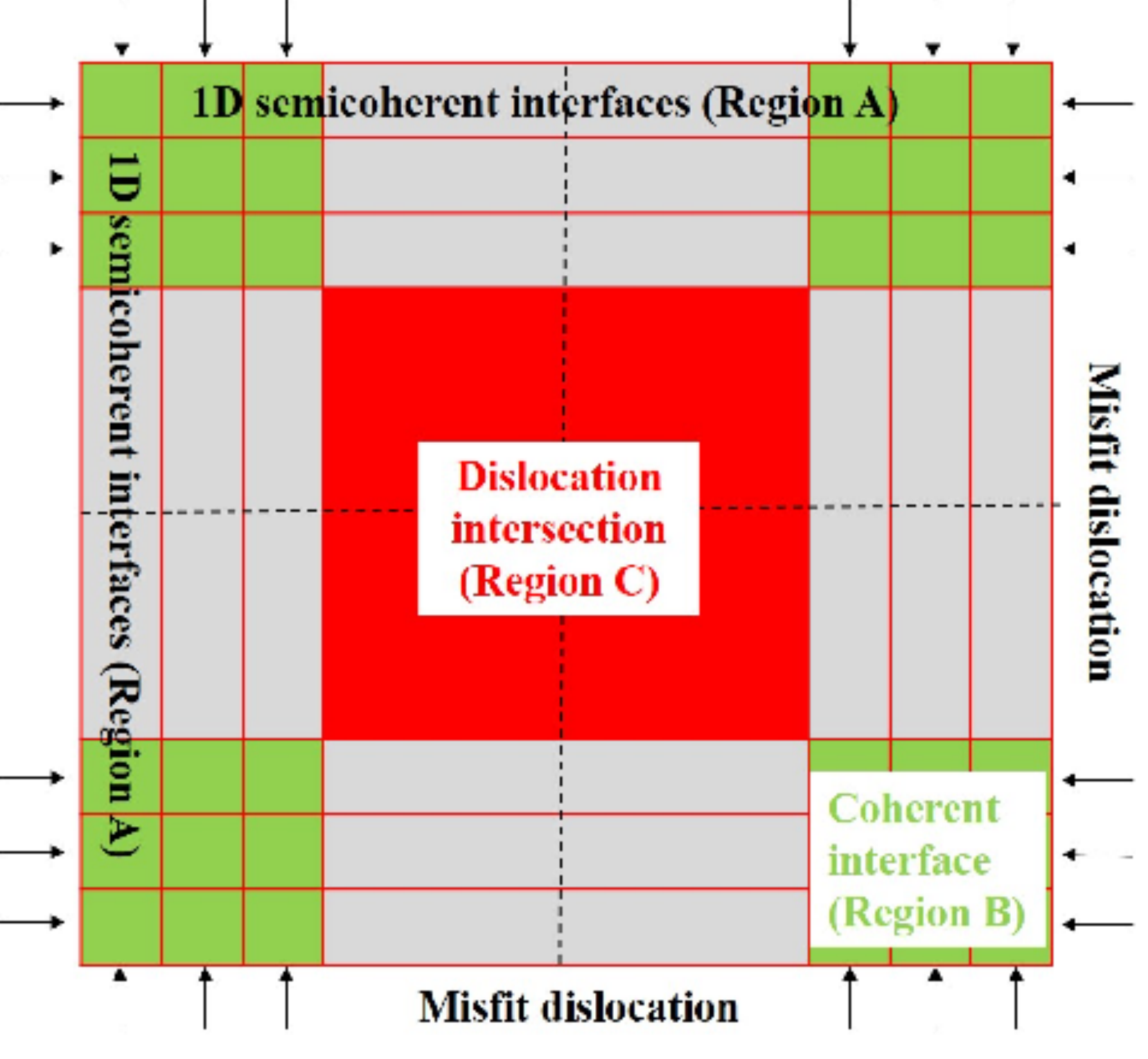} }}%
\qquad
    \subfloat[]{{\includegraphics[width=3cm]{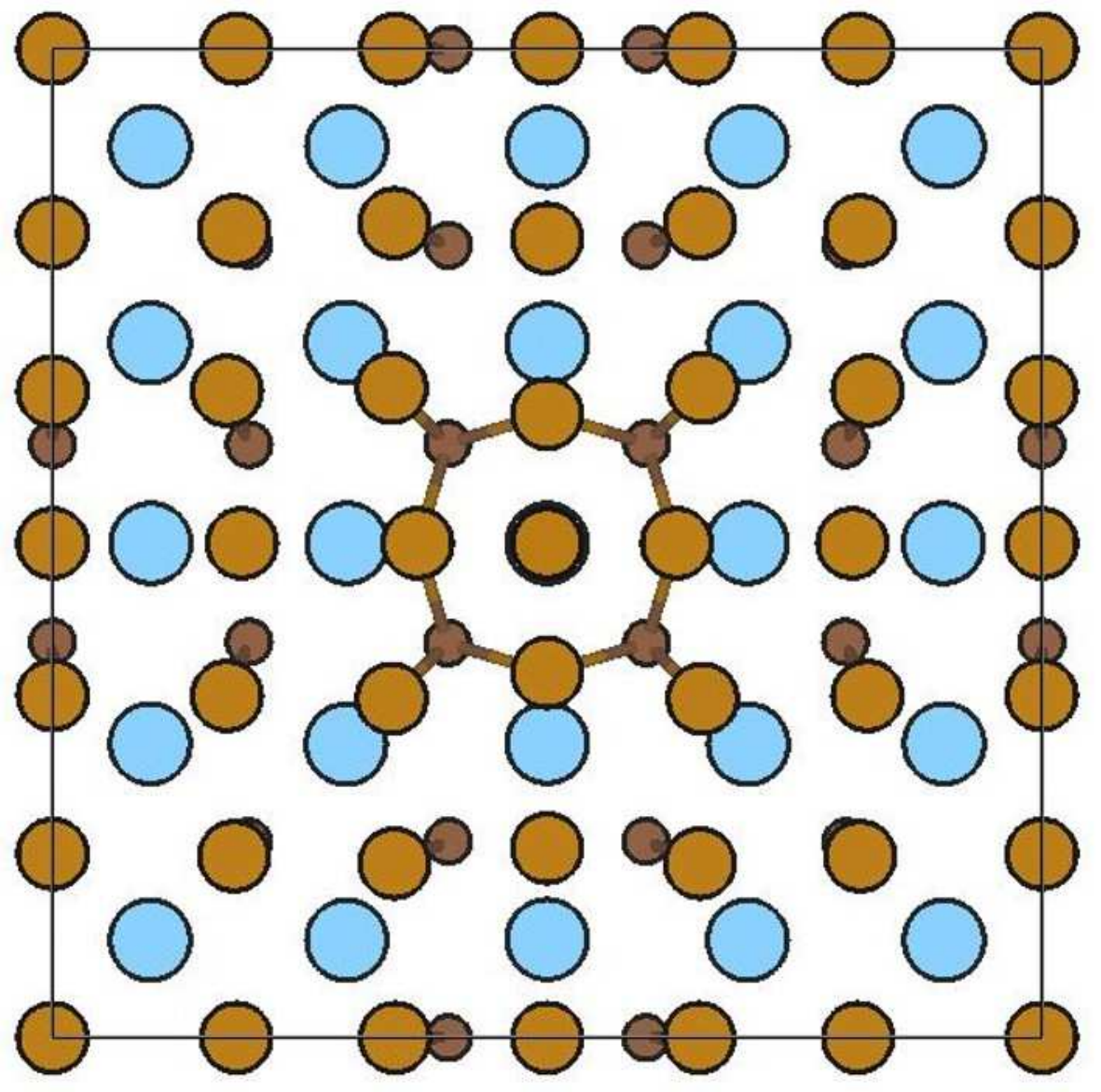} }}%
\qquad
    \subfloat[]{{\includegraphics[width=3cm]{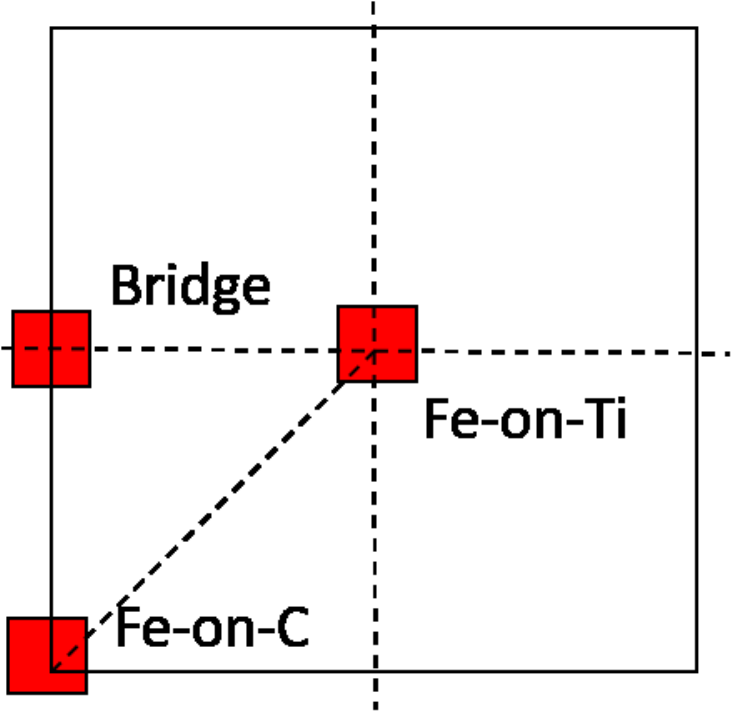} }}%
    \caption{(a) Schematic of the division of semicoherent interface with perpendicular misfit dislocations (dashed lines) into the 1D semicoherent interface slides (Region A, indicated by arrows) and the intersection area of dislocations (Region C, red). Coherent regions (Region B, green) form at the corners. The interface atom configuration after relaxation in the dislocation intersection area (Region C) modeled by 6$\times$6Fe/5$\times$5TiC is shown in (b). (c) Fe vacancy positions at the Fe-on-C, Fe-on-Ti, and Bridge sites at the 6$\times$6Fe/5$\times$5TiC interface. (For interpretation of the references to color in this figure legend, the reader is referred to the Web version of this article.) } %
    \label{semi}%
\end{figure}

Now we apply the same strategy to a more realistic 2D-SI (13$\times$13Fe/12$\times$12TiC) and divide it  into 1D-SIs (region $A$), modeled by the 13$\times$1Fe/12$\times$1TiC ($y=0$) interfaces, and a dislocation core area (region $C$) modeling the intersection part of two perpendicular misfit dislocations. The 2D model is shown in Fig.\ref{semi} (a). In the corners of the 2D-SI where 1D-SI slides cross each other, it is essentially coherent (region $B$). Similarly as in the case of 1D-SI, this division becomes increasingly accurate with larger and larger supercell for the dislocation intersection area ($C$). We compromise between accuracy and computational effort  and thus look for the smallest region $C$ that still leads to acceptable errors. As revealed by our study for the 1D-SIs, the dislocation width decreases when it is close to the dislocation intersection part (Fig.\ref{fig1dstruc} (c)). Therefore, it should be readily enough to adopt a 6$\times$6Fe/5$\times$5TiC  supercell to describe the intersection part. The interface configuration after relaxation is shown in Fig.\ref{semi} (b), from which we can see that it reflects the characteristic patterns as we have expected in Fig.\ref{semi} (a). Notice that the Fe lattice in this  6$\times$6Fe/5$\times$5TiC 2D-SI has been heavily compressed along both $\bf a$ and $\bf b$ directions in order to match the TiC lattice which is held at the equilibrium state. The lattice parameters of the strained tetragonal Fe lattice are 2.55, 2.55, and 3.40 \AA~for $a$, $b$, and $c$, respectively, which should be compared to 2.836 \AA~for the ideal cubic Fe. Its energy is higher than that of equilibrium bcc Fe by 94.75 meV/atom. Using the strained Fe as the reference state, the chemical interfacial energy for the 6$\times$6Fe/5$\times$5TiC 2D-SI  is calculated to be $\sim$0.82 J m$^{-2}$ according to Eq.\ref{eq1}. 

According to the above division, the total interfacial energy for the 2D-SI (13$\times$13Fe/12$\times$12TiC) may be calculated via., 
\begin{equation}\label{eq2de}
\sigma^{\rm est.}_{\rm semi.}=f_A\sigma''_{\rm 1D-13\times1Fe/12\times1TiC}-f_B\sigma''_{\rm coh.}+f_C\sigma''_{\rm 2D-6\times6Fe/5\times5TiC},
\end{equation}
where $f_A$, $f_B$ and $f_C$ are the area fractions of the corresponding regions, calculated as $\frac{168}{144}$, $\frac{49}{144}$ and $\frac{25}{144}$, respectively. The negative sign for the coherent term is to correct the double counting part in $f_A$ (in region $B$). Here we emphasize again that for the real 2D-SI (13$\times$13Fe/12$\times$12TiC), the total interfacial energy that including both chemical and elastic parts should be refereed to the bcc Fe at equilibrium state. Therefore,  the interfacial energies ($\sigma''$) for the 13$\times$1Fe/12$\times$1TiC ($y=0$) 1D-SI, coherent interface, and  the 6$\times$6Fe/5$\times$5TiC 2D-SI in Eq.\ref{eq2de} have been re-calculated with the equilibrium Fe as the reference state to include properly the elastic energy contribution. 
$\sigma''_{\rm 1D-13\times1Fe/12\times1TiC}$, $\sigma''_{\rm coh.}$, and $\sigma''_{\rm 2D-6\times6Fe/5\times5TiC}$ are 0.81, 0.58, and 1.40  J m$^{-2}$, respectively, which are significantly larger than their chemical counterparts (Table 2).  Finally, $\sigma^{\rm est.}_{\rm semi.}$ between ferromagnetic bcc Fe and TiC is calculated to be 0.99 J m$^{-2}$. This value is about 50\% smaller than the previous theoretical value ($\sim$ 1.50 J m$^{-2}$) obtained by Peierls-Nabarro model with the coherent interfacial energy map as input. \cite{Fors2010} The discrepancy is partly due to the effect of relaxation, which has already been disclosed in the case of coherent interface (section 3.1) and also by previous atomic simulation with EAM potential for the Fe/VN semicoherent interface which showed that relaxation decreases the interfacial energy by approximately 50\%, from 0.49 to 0.35 J m$^{-2}$. \cite{Johansson2005, FORS2010550} 

Similarly as the discussion for Eq.\ref{intest1}, $\sigma^{\rm est.}_{\rm semi.}$ in Eq.\ref{eq2de} depends also on the number of Fe layers ($l_{\rm Fe}$). One more Fe layer in the supercells leads to an increment of $\sigma^{\rm est.}_{\rm semi.}$ by $\sim$0.033 J m$^{-2}$. It indicates that on the one hand, $\sigma^{\rm est.}_{\rm semi.}$ is not very sensitive to the number of Fe layers, and on the other hand, one should use the minimum number of Fe layers for the calculations of $\sigma''$ in Eq.\ref{eq2de} to avoid introducing artificial strain energy.

\subsection{Lateral interface}

\begin{figure}[h]%
    \centering
    \subfloat{{\includegraphics[width=8cm]{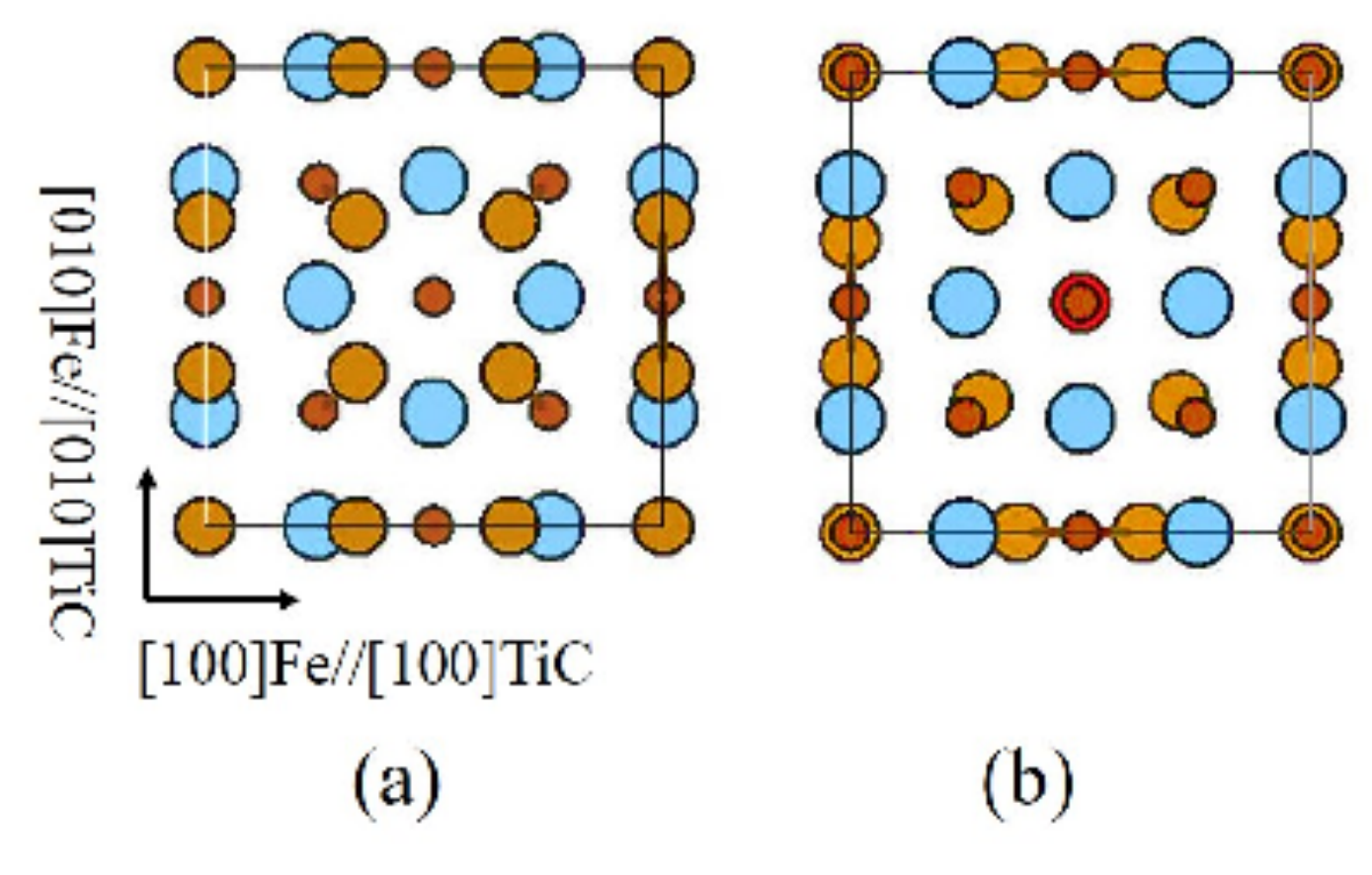} }}%
    \caption { Schematic of the commensurate interface model for the lateral interface, (a) before relaxation and (b) after relaxation. Only the atoms in the interface Fe and TiC layers are shown, except that the  Fe atom (red) in the center of (b) which originally belongs to the second Fe layer away from the interface relaxes to the interface layer. (For interpretation of the references to color in this figure legend, the reader is referred to the Web version of this article.)} %
    \label{com}%
\end{figure}

Wei et al. revealed that the TiC particles show a moderate tendency to form faceted lateral interfaces.~\cite{wei2004} Taniguchi et al. observed that TiC precipitates were in the shape of plate-like rectangular parallelepipeds using aberration-corrected scanning TEM. \cite{Taniguchi2012} While the exact OR for the lateral interface is not clear, based on geometrical consideration of the atomic arrangement between the two phases related by the B-N OR at the broad interface and HRTEM images, periodic structures may exists along the [100]$_{\rm Fe}$ and [100]$_{\rm TiC}$ directions in the lateral interface. \cite{wei2004} Here we choose the orientation relationship, (100)Fe//(100)TiC and [100]Fe//[100]TiC, for the representation of the lateral interface, (Fig.\ref{com}). Following this orientation relationship, a commensurate interface structure composed of 3$\times$3 Fe units meeting 2$\times$2 TiC units possesses very small lattice mismatches. The lattice mismatches calculated with the theoretical and experimental lattice parameters are 1.79\% and 0.88\%, respectively. We notice that other ORs for the lateral interface, i.e., (110)$_{\rm Fe}$//(001)$_{\rm TiC}$ or (010)$_{\rm Fe}$//(011)$_{\rm TiC}$, may also exist. \cite{Yang2002184, Arya2003} However, the calculated coherent interfacial energy of the (110)$_{\rm Fe}$//(001)$_{\rm TiC}$ interface is very close to that of (001)$_{\rm Fe}$//(001)$_{\rm TiC}$, which  cannot explain the observed plate-like morphology at the  early stage of nucleation when the precipitates are fully coherent with matrix.~\cite{KOBAYASHI2012854, MUKHERJEE20132521, MUKHERJEE2017621} Further more, the theoretical length of lattice vector $\bf{a}$ for above commensurate interface is 8.66 \AA~, which is consistent with the atom probe tomography (APT) results showing that the TiC platelets at the nucleation stage are usually very thin ($<$1 nm) (which are obtained after aging at low temperatures). \cite{Takahashi2010261}

Similar as for the other interfaces considered previously, when taking the correspondingly strained Fe and equilibrium TiC lattices as reference, the calculated chemical interfacial energy for the lateral interface is $\sim$1.77 J m$^{-2}$. When taking equilibrium Fe as reference, the interfacial energy is 1.80 J m$^{-2}$. The latter is only about 1.7\% larger than the chemical part, which is consistent with the small lattice mismatches at the commensurate interface.

\begin{table}[tb!]
\caption{The calculated interfacial energies (J m$^{-2}$) for the 1D semicoherent and coherent interfaces with different reference energies for Fe.  The chemical interfacial energies ($\sigma^{\rm chem.}$) are calculated referring to the Fe lattice having the same strain state as in the the corresponding interface supercell. For the calculations of $\sigma'$, and $\sigma''$, the reference energies for Fe are   $E_{\rm Fe}^{\rm 13\times1Fe/12\times1TiC}$ and $E_{\rm Fe}^{\rm eq.}$, respectively. The lattice parameters ($a$, $b$, $c$, \AA) for the strained Fe lattices in the coherent and 1D semicoherent interfaces, and the corresponding energy difference ($\Delta E_{\rm Fe}$, meV/atom) relative to that of equilibrium bcc Fe ($E_{\rm Fe}^{\rm eq.}$) are also listed in the lower panel of the Table.}
\label{chem}
\smallskip
\small
\begin{threeparttable}
\begin{tabular}{llllllll}
\hline\hline
& \multicolumn{2}{c}{13$\times$1Fe/12$\times$1TiC} &\multicolumn{2}{c}{6$\times$1Fe/5$\times$1TiC} &\multicolumn{3}{c}{Coh. ( $x=0$)}\\
\cline{2-3}\cline{4-5}\cline{6-8}
$y$ & $\sigma^{\rm chem.}$ &$\sigma''$  & $\sigma^{\rm chem.}$ &$\sigma'$ & $\sigma^{\rm chem.}$ &$\sigma'$  &$\sigma''$  \\
\hline
0   & 0.65&0.81& 0.75 &1.03&0.22&0.44&0.58 \\
0.1&0.87&1.02& 0.98&1.25&0.42&0.64&0.78 \\
0.2& 1.40&1.56& 1.55 &1.83&0.90&1.12&1.26\\
0.3& 1.98&2.13&2.20  &2.47&1.38&1.60&1.74\\
0.4& 2.36&2.51&2.64 &2.91&1.69&1.91&2.05\\
0.5& 2.47&2.62& 2.77 &3.04&1.78&2.00&2.14\\
\hline
$a$&2.827  &&2.552 &&3.062&&\\
$b$&3.062  &&3.062 &&3.062&&\\
$c$&2.740  &&3.000 &&2.660&&\\
$\Delta E_{\rm Fe}$&32.95  &&86.20&& 84.70&&\\
\hline\hline
\end{tabular}
\end{threeparttable}
\end{table}

\section{Discussion}

\subsection{Aspect ratio of the disc-like TiC precipitates}

Morphology is an important factor that affects TiC particle strengthening in steels. When considering the particle  size-dependent properties, it is not sufficient using the spherical particle assumption. For example, the size of TiC particles in ferrite is usually expressed as an equivalent volume diameter of a spherical particle in the experimental results. \cite{KOBAYASHI2012854, MUKHERJEE20132521, MUKHERJEE2017621} Therefore, a particle with small equivalent diameter does not necessarily mean that it forms coherent interface with matrix because of its disc-like shape as observed. It causes difficulty in understanding, for example, the particle size dependence of the interaction force between dislocation and TiC precipitate. \cite{KOBAYASHI2012854}

\begin{figure}[h]%
    \centering
    \subfloat[]{{\includegraphics[width=8cm]{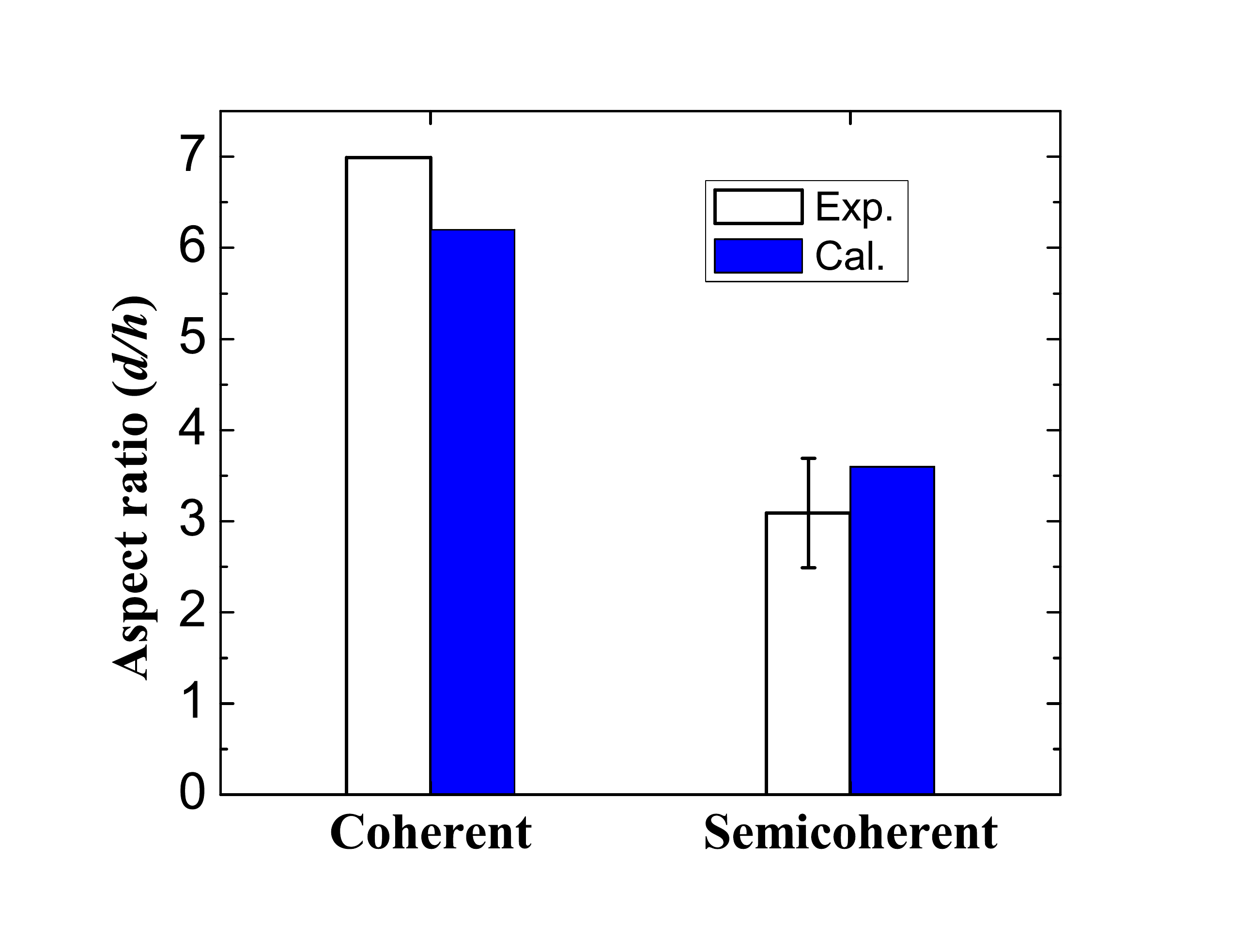} }}%
    \caption{The calculated aspect ratio ($d/h$) for TiC particles with coherent and semicoherent types of broad interfaces, respectively. The experimental aspect ratios correspond to small and large TiC precipitates obtained at low and high tempering temperatures, respectively. \cite{Wei2006} } %
    \label{aspect}%
\end{figure}

Considering a disc-like shape precipitate with diameter $d$ and height $h$, the equilibrium aspect ratio ($d/h$) is decided by the balance of energies of the broad and lateral interfaces via.,
\begin{equation}
2\pi(\frac{d}{2})^2\sigma_1=\pi dh\sigma_2,
\label{aspecteq}
\end{equation}
where $\sigma_1$ and $\sigma_2$ are the interfacial energies for the broad and lateral interfaces, respectively. From the above equation, we arrive at $d/h=2\sigma_2/\sigma_1$. Using the obtained interfacial energies, $\sigma_1=0.58$ and $\sigma_1=0.99$  J m$^{-2}$, for the coherent and semicoherent broad interfaces, and $\sigma_2=1.80$ J m$^{-2}$ for the lateral interface, $d/h$ ratios are calculated to be 6.2 and 3.6, respectively. It is in a nice agreement with the observed shape evolution of TiC particles with increasing size, Fig.\ref{aspect}, \cite{Wei2006, Takahashi2010261} which should be interpreted as an indirect evidence for the chosen OR for the lateral interface and also supports our calculated interfacial energy and specifically the interface energy anisotropy for the Fe/TiC interface.

\begin{figure*}[tbh!]%
 \centering
 \subfloat{{\includegraphics[width=18cm]{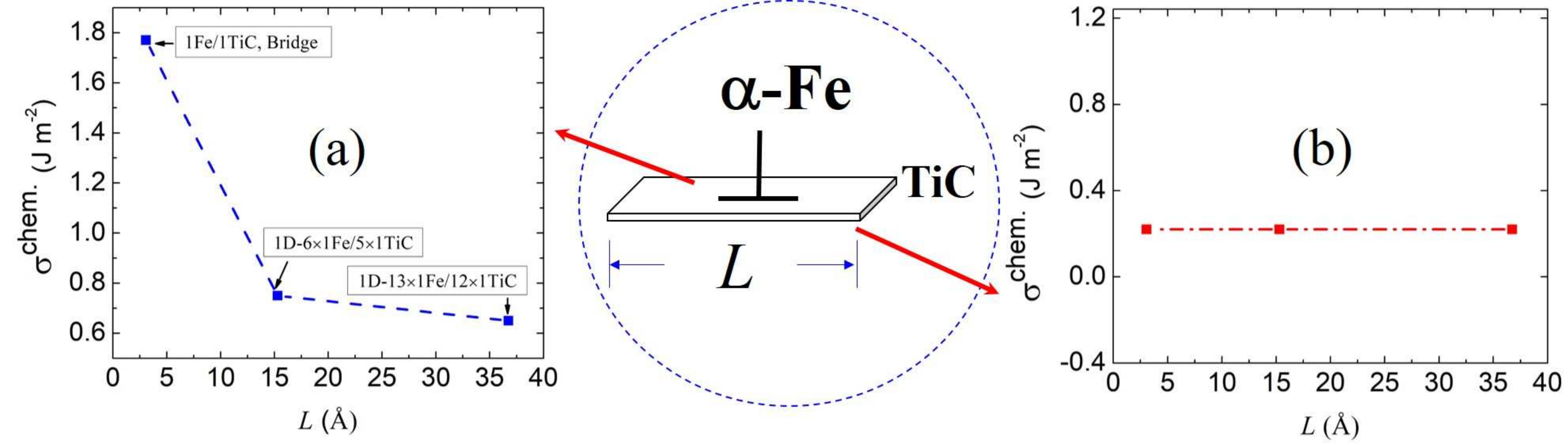}}}%
\caption{The size dependence of the chemical interfacial energy for the upper (a) and lower (b) facets of a strip-like TiC precipitate nucleating on an edge dislocation, respectively.  The upper and lower facets of the precipitate are in contact with the compressive and tensive stain fields of the edge dislocation, respectively. They correspond to the 1D semicoherent and coherent interfaces, respectively.}%
\label{fignm}%
\end{figure*}

\subsection{Energetics for precipitate nucleation on dislocation}

Experimentally, the radius of interphase VC nucleus is found to be approximately in the range of 0.2-1.0 nm, assuming a spherical shape with semicoherent interface. \cite{ZHANG2017166} The observed VC particles are commonly less than 2 nm. After transforming to a disc-like shape using aspect ratio $d/h=10$, \cite{KOBAYASHI2012854} $d$ is found to be much smaller than the theoretical periodicity of misfit dislocations ($\sim$8 nm). Usually, it indicates coherent interface. However, the experimental interfacial energy\cite{ZHANG2017166} is $\sim$2-4 times larger than the calculated semicoherent interfacial energy ($\sim$0.55 J m$^{-2}$\cite{Fors2010}). One reason for the discrepancy may be ascribed to the oversimplified model for nucleation which commonly assume spherical shape of nucleus and a constant interfacial energy. \cite{ZHANG2017166, Jang2013} Another reason may be related to the lack of detailed information about the interface structures, for example, interfacial vacancies are ignored in both experimental and theoretical consideration of interfaces. \cite{Harford2000} In general,  it is known that vacancy concentrates to dislocation core or grain boundary \cite{Bullough1970101} and is a constitutional type of defect in transition metal carbides/nitrides.~\cite{Jang2012208}  In the following, we attempt to  discuss some of these aspects.

 It is commonly observed that metal carbides and nitrides prefer nucleation on dislocations. Carbon and nitrogen tend to segregate around dislocations and form the so-called ``Cottrell atmosphere''. It was showed that  the saturate C concentration near dislocation core reaches as high as $\sim$10 at.\%. \cite{WILDE200039} For an edge dislocation, at low carbon concentration, C atoms tend to occupy interstitial sites within the traction zone , whereas at higher C concentration, it may also occupy the sites in the compression zone. This asymmetric atmosphere of C leads to slightly relaxed stress in the compression zone and to higher tensile stress in the traction zone. \cite{WASEDA201716} Considering the strong affinity between transition metals with  C and N and the high pipe-diffusion rate at dislocation core~\cite{Legros1646}, it is not surprising that substitutional atoms are also drawn to the expansive zone and form the nucleation of MX precipitate. \cite{KWIATKOWSKIDASILVA2017305, TAKAHASHI2016415, DANG2007557} Furthermore, in general near an edge dislocation core larger substitutional atoms (Ti, V, Nb, etc.) than Fe are expected to occupy sites with the expansive strain. \cite{TAKAHASHI2016415} Therefore, the nucleation process is more decided by the diffusion of transition metal elements (V, Ti, etc.) than that of interstitials (C, N). It is consistent with the experimental observation that the density of VC interphase precipitates increases with V content, while it barely changes with higher C content. \cite{ZHANG2017166,ZHANG2015375}

In APT experiments, it has been demonstrated that in the very early stage of nucleation, particles or clusters of MX are also disc-like, ranging from 1 to several atomic layers in thickness and following the B-N OR.~\cite{KOBAYASHI2012854, MUKHERJEE20132521, MUKHERJEE2017621,Takahashi2010261} Since there are  no clear experimental evidence for the the degree of coherency for the interfaces when nucleation is on dislocations, here we propose two possible models for the nucleation process assisted by edge dislocation and discuss how the interfacial energy changes with respect to precipitate size.

\subsubsection{Nucleation inside the expansive zone}
Even though the size of the expansive zone created by edge dislocation is not well measured, the radius of Cottrell atmosphere of C was measured as large as $\sim$7 nm. \cite{WILDE200039} It is therefore possible for a TiC nucleus to be fully embedded in the expansive zone. In this case, approximately only the chemical part of coherent interfacial energy ($\sigma_{\rm coh.}^{\rm chem.}=0.22 $ J m$^{-2}$) determines the nucleation barrier, because the local Fe lattice has already been expanded by the existence of the edge dislocation and C interstitials. Therefore, this nucleation process is much easier than the homogeneous nucleation which needs to create chemical interfaces between precipitate and matrix,  and at the same time, the surrounding coherent strain field as well. We also expect a larger aspect ratio ($d/h\approx16$) according to Eq.\ref{aspecteq}. 

\subsubsection{Nucleation at the interface between the expansive and compressive zones}

Since C concentration decays from the dislocation core, \cite{WASEDA201716} the TiC nucleus may be in contact with both of the compressive and expansive zones that are created by an edge dislocation.  The existence of the stain field around the dislocation core facilitates the nucleation of precipitate by forming coherent interface on the side facing the expansive zone and semicoherent interface on the other side towards the compressive zone, (Fig.\ref{fignm}). In this case, energetically, it should be more favorable because forming coherency on both sides involves the displacement of the original dislocation segment away from the newly formed coherent interface and the reconstruction of elastic field. It is expected that the TiC nucleus has a strip-like shape and firstly grows in the direction perpendicular to dislocation line, which gradually decreases  the chemical interfacial energy until the formation of a new misfit dislocation. The dependence of chemical interfacial energy on the length of TiC nucleus for the interfaces in contact with the compressive and expansive zones are plotted in Fig.\ref{fignm} (a) and (b), respectively. After nucleation, the particles can then grow along dislocation line  or perpendicular to it, with the interfaces gradually changing to 2D semicoherent interface with dislocation network. Even though some strip-like clusters of TiC with less than 50 atoms can be easily identified in the atom probe microscopy results,~\cite{MUKHERJEE20132521, MUKHERJEE2017621} which is consistent with the above scenario, more careful studies are required to fully support the proposed models. 

\subsection{Formation of vacancies at the semicoherent interface}

\begin{table}[tb!]
\caption{Vacancy formation energies ($\Delta E_v$, eV/atom) at the Fe-on-C, Bridge, and Fe-on-Ti sites at the interface Fe layer in the 2D 6$\times$6Fe/5$\times$5TiC semicoherent interface, according to Eq.\ref{vac}. In ferromagnetic bulk $\alpha$-Fe, the vacancy formation energy is calculated with the equilibrium Fe as the reference. Previous theoretical and experimental results in bulk $\alpha$-Fe are also listed for comparison.}
\label{vac}
\smallskip
\small
\begin{threeparttable}
\begin{tabular}{llllllc}
\hline\hline
\multicolumn{3}{c}{2D semi. interface}& & Bulk Fe\\
\cline{1-3}     \cline{5-6}
  Fe-on-C & Fe-on-Ti & Bridge&  & Exp.&Cal.\\
\hline
2.12 & -1.00 & 0.51 & & 1.4$\pm$0.1\tnote{b}&2.16\\
 & & & &2.0$\pm$0.2\tnote{c} &2.25\tnote{a}, 2.32\tnote{d} \\
\hline\hline
\end{tabular}
\begin{tablenotes}
\item[a]Ref.~\cite{Korzhavyi1999}
\item[b]Ref. \cite{0305-4608-8-5-001}
\item[c]Ref.~\cite{De1983}
\item[d]Ref.~\cite{Delange2016}
\end{tablenotes}
\end{threeparttable}
\end{table}

In general, dislocations are known as efficient vacancy sinks, and abundant vacancies at dislocation core assist the fast pipe diffusion of impurities. \cite{Legros1646} At semicoherent interfaces, in the vicinity of misfit dislocation cores and dislocation intersections, we have observed significant local lattice distortion, especially in the Fe lattice, e.g., in Fig.\ref{fig1dstruc} (a), (b) and Fig.\ref{semi} (b). Misfit dislocations are therefore expected to have a great impact on the behaviors of defects or alloying segregation at the interface. Here, as an example, we have calculated the vacancy formation energies ($\Delta E_{v}$) for three sites (Fig.\ref{semi} (c)) in the interface Fe layer in the 6$\times$6Fe/5$\times$5TiC  2D-SI. The results are listed in Table \ref{vac}. Locally, three sites have the Fe-on-C, Fe-on-Ti, and Bridge configurations, respectively. The vacancy formation energy is calculated with respect to the strained Fe which is under the same strain state as in the 6$\times$6Fe/5$\times$5TiC 2D-SI, via.,
\begin{equation}\label{vac}
\Delta E_v (i)=E^{vac_i}_{\rm Fe/TiC}+E_{\rm Fe}^{\rm 6\times6Fe/5\times5TiC}-E_{\rm Fe/TiC},
\end{equation}
where $E^{vac_i}_{\rm Fe/TiC}$ and $E_{\rm Fe/TiC}$ are the total energies for the 6$\times$6Fe/5$\times$5TiC 2D-SIs with and without a vacancy (i) at the interface Fe layer, respectively. $E_{\rm Fe}^{\rm 6\times6Fe/5\times5TiC}$ is the reference energy for Fe in the 2D-SI.
 Our results show that at the coherent region (Fe-on-C), the vacancy formation energy is only slightly smaller than that in bulk Fe. While at the dislocation core (Bridge), $\Delta E_v$ is decreased by more than 75\% compared to that in bulk Fe. At the position where dislocations cross each other (Fe-on-Ti), $\Delta E_v$ is even negative which indicates that vacancy is an intrinsic defect at the semicoherent interface. On the one hand, our results show that the formation of vacancy at misfit dislocation core is much easier that in bulk material, on the other hand, it indicates a different local atomic structure (decorated by vacancies) for the intersection region of dislocations from what one would normally expect. \cite{Davide2016, Jung2010} The presence of intrinsic vacancies at the interface has deep implication on many aspects, such as alloying segregation, pipe diffusion, and  H trapping \cite{Davide2016, Chen1196}, etc.




\section{Summary}
We have taken the Fe(001)/TiC(001) semicoherent interface as an example to explore the way for estimating the interfacial dislocation structures and energetics. We demonstrate how one can rationally divide the semicoherent interface  into small patches that fit ab initio calculations, based on our studies for the so-called 1D semicoherent interface.  Even for the interface system with very small lattice mismatch, e.g., VC/Fe with $\sim$2\% mismatch, the 1D-SI is still within the capability of modern ab initio methods. Direct ab initio studies for the 1D-SIs produce valuable information about misfit dislocations, for example, the spatial  extension of strains and the size of dislocation core, etc., which guides to properly perform the division.  The present method takes into account both chemical and elastic parts of the interfacial energy in the semicoherent interfaces with square dislocation network. Importantly, it can be adopted to study the effects of vacancy and alloying on the interfacial energy, which are missing in the previous Peierls-Nabarro framework.

\section*{Acknowledgment}

We are grateful to Hans Magnusson, Sten Wessman (Swerea KIMAB), Mattias Thuvander (CTH), Qing Chen (Thermo-Calc Software), Jan Y Jonsson (Outokumpu Stainless), Magnus Andersson, Ulrika Borggren (SSAB), Claes Olsson, Lars Nyl{\o}f (Sandvik Materials Technology), Sebastian Ejnermark (Uddeholm), and Rachel Pettersson (Jernkontoret) for helpful discussions. The  present work is performed under the project ''Future Materials Design'' financed  by the Swedish steel producers' association (Jernkontoret) and the Sweden's innovation agency (Vinnova). Financial support by the Swedish Research Council, the Swedish Foundation for Strategic Research, the Swedish Foundation for International Cooperation in Research and Higher Education, and the Hungarian Scientific Research Fund (OTKA 109570) are also acknowledged. The Swedish National Supercomputer Centre  (NSC) and the High Performance Computing Center North (HPC2N)  are acknowledged for providing computation resources.

\section*{Reference}
\balance
\bibliographystyle{model1-num-names}
\bibliography{ref}
\end{document}